# Exceptional damage-tolerance of a medium-entropy alloy CrCoNi at cryogenic temperatures

Bernd Gludovatz[1], Anton Hohenwarter[2], Keli V.S. Thurston[1,3], Hongbin Bei[4], Zhenggang Wu[5], Easo P. George[4,5,†] & Robert O. Ritchie[1,3]

High-entropy alloys are an intriguing new class of metallic materials that derive their properties from being multi-element systems that can crystallize as a single phase, despite containing high concentrations of five or more elements with different crystal structures. Here we examine an equiatomic medium-entropy alloy containing only three elements, CrCoNi, as a single-phase face-centred cubic solid solution, which displays strength-toughness properties that exceed those of all high-entropy alloys and most multi-phase alloys. At room temperature, the alloy shows tensile strengths of almost 1 GPa, failure strains of ~70% and $K_{JIc}$ fracture-toughness values above 200 MPa m$^{1/2}$; at cryogenic temperatures strength, ductility and toughness of the CrCoNi alloy improve to strength levels above 1.3 GPa, failure strains up to 90% and $K_{JIc}$ values of 275 MPa m$^{1/2}$. Such properties appear to result from continuous steady strain hardening, which acts to suppress plastic instability, resulting from pronounced dislocation activity and deformation-induced nano-twinning.

[1] Materials Sciences Division, Lawrence Berkeley National Laboratory, Berkeley, California 94720, USA. [2] Department of Materials Physics, Montanuniversität Leoben and Erich Schmid Institute of Materials Science, Austrian Academy of Sciences, Leoben 8700, Austria. [3] Department of Materials Science and Engineering, University of California, Berkeley, California 94720, USA. [4] Materials Sciences and Technology Division, Oak Ridge National Laboratory, Oak Ridge, Tennessee 37831, USA. [5] Department of Materials Sciences and Engineering, University of Tennessee, Knoxville, Tennessee 37996, USA. † Present address: Institute for Materials, Ruhr University, 44801 Bochum, Germany. Correspondence and requests for materials should be addressed to E.P.G. (email: easo.george@rub.de) or to R.O.R. (email: roritchie@lbl.gov).

Equiatomic multi-component metallic materials, referred to variously as high-entropy alloys (HEAs), multi-component alloys or compositionally complex alloys, have generated considerable excitement in the materials science community of late as a new class of materials that derive their properties not from a single dominant constituent, such as iron in steels, but rather from multiple principal elements with the potential for unique combinations of mechanical properties compared with conventional alloys[1–19]. Much of the interest is predicated on the belief that many new alloys with useful properties are likely to be discovered near the centres (as opposed to the corners) of phase diagrams in compositionally complex systems[17].

One of the extensively investigated high-entropy alloys, an equiatomic, face-centred cubic (fcc) metallic alloy comprising five transition elements, Cr, Mn Fe, Co and Ni, was introduced in 2004 (ref. 1), although it was not for a decade that its mechanical properties were first systematically characterized[6–8,13–15]. CrMnFeCoNi (often termed the Cantor alloy[1]) displays strongly temperature-dependent strength and ductility with only a small strain-rate dependence[6,7]. Furthermore, between room temperature and 77 K, the alloy displays fracture toughness, $K_{JIc}$, values at crack initiation that remain well above 200 MPa m$^{1/2}$ associated with an increase in tensile strength (763 → 1,280 MPa) and ductility (0.5 → 0.7), making it not simply an ideal material for cryogenic applications but putting it among the most damage-tolerant materials in that temperature range[8].

Although this excellent combination of properties can be related to progressively increasing strain hardening with hardening exponents above 0.4 (refs 7,8), it remains unclear why this particular combination of elements with very different crystal structures produces a single-phase microstructure[7,10–12,20,21], whereas many others with comparable configurational entropies do not[5]. In fact, a relatively small number of the reported multi-element high-entropy alloys are simple solid solutions[22]. Cantor et al.[1] produced an alloy with 20 elements in equal atomic ratios that crystallized as a very brittle multi-phase microstructure indicating that high configurational entropy by itself is unable to suppress the formation of intermetallic phases comprising the constituent elements. As pointed out recently[16], while the equiatomic composition maximizes configurational entropy, it does not necessarily minimize the total Gibbs free energy of a multi-component solid solution, and increasing the number of constituent elements could actually lead to the formation of undesirable intermetallic phases. Clearly, it is the nature of the alloying elements and not just their sheer number that is relevant.

However, there is also a question of the role of high configurational entropy in these materials with respect to properties, particularly how such high-entropy alloys compare with other equiatomic multi-element systems. Here we examine a variant of the single-phase CrMnFeCoNi high-entropy alloy in which two of the elements have been removed. The resulting CrCoNi alloy, an equiatomic 'medium-entropy alloy' (MEA), has a single-phase, fcc crystal structure[23], whose uniaxial tensile properties have recently been reported[24]. The experimental results (X-ray diffraction and backscattered electron, BSE, images)[23] are consistent with the CrCoNi ternary phase diagrams[25], which indicate that the equiatomic composition is a single-phase solid solution at elevated temperatures. (XRD and BSE analyses of the five-component CrMnFeCoNi alloy after casting and homogenization[5] showed that it too is single-phase fcc and remains so after recrystallization when examined by transmission electron microscopy[7]. In addition, three-dimensional atom probe tomography on the five-component CrMnFeCoNi alloy in the cast/homogenized state[21] and after severe plastic deformation[20] have shown that it retains its true single-phase character at the much finer atomic scale.

Significantly, we find here that the fracture toughness properties of the three-component CrCoNi MEA are even better than those of the five-component CrMnFeCoNi HEA, and are further enhanced with decrease in temperature between 293 and 77 K, making it one of the toughest metallic materials reported to date.

## Results

**Microstructure.** The CrCoNi MEA was produced from high-purity elements (>99.9% pure) which were arc-melted under argon atmosphere and drop-cast into rectangular cross-section copper moulds followed by cold forging and cross rolling at room temperature into sheets of roughly 10 mm thickness (Fig. 1a). Following recrystallization, optical microscopy (Fig. 1b), scanning electron microscopy (SEM) (Fig. 1c) and electron back-scattered diffraction (EBSD; Fig. 1d) images taken from the cross-section of the sheets revealed an equiaxed grain structure with a variable grain size of 5–50 μm and numerous recrystallization twins (inset of Fig. 1c); the equiatomic elemental distribution of the alloy can be seen from energy-dispersive X-ray (EDX) spectroscopy in Fig. 1e. Uniaxial tensile specimens and compact-tension C(T) fracture-toughness specimens were cut from the sheets using electrical discharge machining; the C(T) samples were fatigue precracked and subsequently side-grooved, in general accordance with ASTM standard E1820 (ref. 26).

**Strength and ductility.** Using uniaxial, dog-bone-shaped tensile specimen, we measured stress–strain curves at room temperature (293 K), in a mixture of dry ice and ethanol (198 K), and in liquid nitrogen (77 K). Results in Fig. 2a show a ∼50% increase in both yield strength, $\sigma_y$, and ultimate tensile strength, $\sigma_{UTS}$, with decreasing temperature to values of $\sigma_y = 657$ MPa and $\sigma_{UTS} = 1,311$ MPa at 77 K. The tensile ductility (strain to failure, $\varepsilon_f$) similarly increased by ∼25% to ∼0.9, leading to an increase in fracture energy of more than 80%, associated with a high strain-hardening exponent, $n$ of 0.4. (Note that compared with pure Ni, this material displays both higher strain hardening and higher elongation to failure[24], consistent with the widely accepted Considère's criterion that higher work-hardening ability promotes ductility by postponing plastic (geometric) instability.)

The yield strength of this alloy is not particularly high, although it does significantly strain harden to give low-temperature tensile strengths above 1 GPa. However, as discussed below, its outstanding characteristic is a combination of high strength, ductility and especially fracture toughness which is enhanced significantly at cryogenic temperatures. This refers to its damage tolerance which is invariably the most important property for the application of a structural material.

**Fracture toughness.** To assess the fracture toughness of the CrCoNi alloy and account for both the elastic and extensive plastic contributions involved in the deformation process and during crack growth, we applied nonlinear-elastic fracture mechanics analysis to determine $J$-based crack-resistance curves, that is, $J_R$ as a function of crack extension $\Delta a$, as shown in Fig. 2b. At room temperature, our C(T) specimens show fracture toughness, $J$, values in excess of 200 kJ m$^{-2}$ at crack initiation, which increased to above 400 kJ m$^{-2}$ with crack extensions of slightly more than 2 mm, the maximum extent of cracking permitted for this geometry by ASTM standards[26]. Despite the much higher strength at lower temperatures, at 77 K the critical $J$ increased even further to above 350 kJ m$^{-2}$ at crack initiation and to almost 950 kJ m$^{-2}$ at full extension of the crack. Given that the requirements for $J$-dominant conditions, that is, $b, B >> 10 \ (J\ \mathrm{per}\ \sigma_{flow})$, where $b$ is the uncracked ligament width (sample width, $W - a$), $B$ the sample thickness and $\sigma_{flow}$ the flow stress ($\sigma_{flow} = (\sigma_y + \sigma_{UTS})/2$), were met, the standard J–K equivalence (mode I) relationship, $K_J = (J\ E')^{1/2}$, was used to determine stress-intensity $K$ values corresponding to these measured $J$ toughnesses. (Here, $E' = E$, the Young's modulus in plane stress and $E/(1-\nu^2)$ in plane strain, $\nu$ is the Poisson's ratio, where values of $E$ and $\nu$ were determined at each temperature using resonance ultrasound spectroscopy methods described elsewhere[27].) Fracture toughnesses for the CrCoNi alloy, defined at crack initiation, were strictly valid by ASTM Standard E1820 (ref. 26), with measured $K_{JIc}$, values of 208 MPa m$^{1/2}$ ($J_{Ic} = 212$ kJ m$^{-2}$) at 293 K increasing to 273 MPa m$^{1/2}$ ($J_{Ic} = 363$ kJ m$^{-2}$) at 77 K. ASTM valid crack-growth toughnesses, defined at $\Delta a \sim 2$ mm, were significantly higher with critical stress-intensity values above $\sim 290$ MPa m$^{1/2}$ ($J \sim 400$ kJ m$^{-2}$) at 293 K, rising up to $\sim 430$ MPa m$^{1/2}$ ($J \sim 900$ kJ m$^{-2}$) at 77 K.

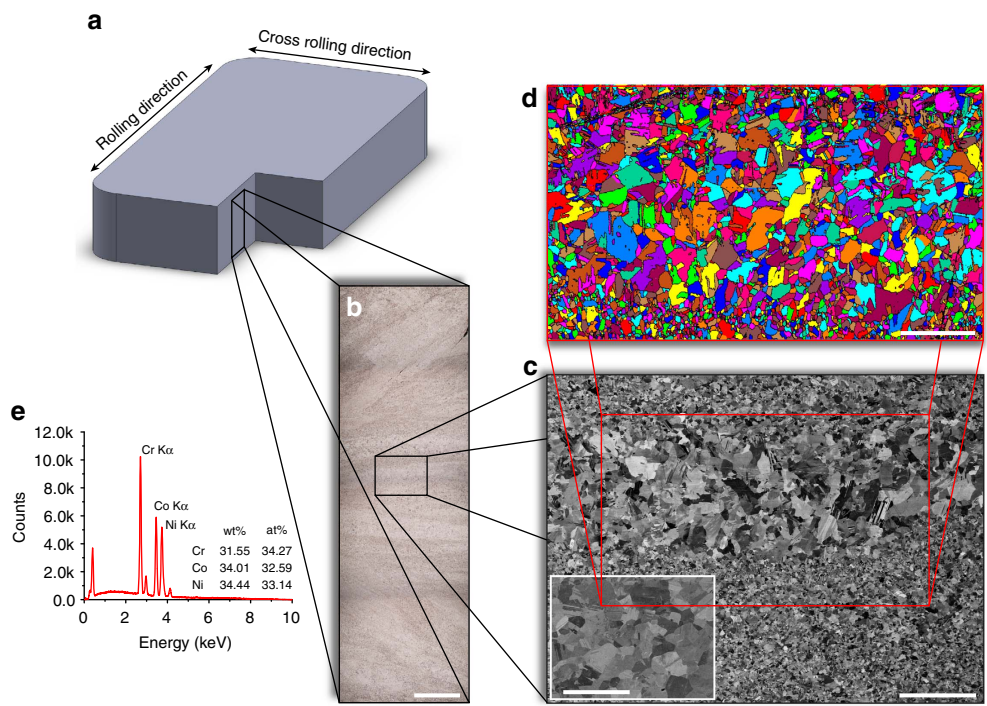

**Figure 1 | Processing and microstructure of the medium-entropy alloy CrCoNi.** (**a**) The material was processed by arc melting, drop casting, forging and rolling into sheets of roughly 10 mm thickness from which samples for cross-sectional analysis, tensile tests and fracture toughness tests were machined. (**b**) Optical microscopy image shows the varying degree of deformation through the thickness of the sheets. (**c**) Scanning electron microscopy images reveal the non-uniform grain size of the material resulting from the deformation gradients, equiaxed grains and numerous annealing twins after recrystallization (inset). (**d**) Grain maps from electron back-scatter diffraction scans confirm the varying grain size and show the fully recrystallized microstructure. (**e**) Energy-dispersive X-ray spectroscopy verifies the equiatomic character of the alloy. The scale bars in **b**,**c** and the inset of **c** and **d** are 1 mm, 200 μm, 20 μm and 150 μm, respectively.

**Deformation and failure mechanisms**. The extremely high fracture toughness values of the CrCoNi alloy (Fig. 2) are associated with fully ductile fracture, with a pronounced stretch-zone at crack initiation (Fig. 2c) and failure by microvoid coalescence (Fig. 2d). The volume fraction of the void-initiating inclusions was lower than in the five-component CrMnFeCoNi alloy[8], which is partly an effect of removing Mn that is known to increase the number of inclusions[6]. The particles here were analysed by EDX spectroscopy and found to be Cr-rich (insets of Fig. 2d), whereas in the five-component HEA, both Cr and Mn-rich particles were found[8]. In both alloys, we believe that these particles are oxide inclusions that typically form when alloys containing reactive elements are melted. Consistent with this, a recent study identified $MnCr_2O_4$ oxide particles in an induction melted CrMnFeCoNi HEA by EDX analysis[28]. To quantify their effects on ductility and fracture, further studies are needed in the future that would accurately determine the relative volume fractions of the void-initiating inclusions in the MEA and HEA and identify their chemistry and crystal structure by TEM after extraction from the voids.

While the yield strength and $K_{JIc}$ fracture toughness of the medium-entropy CrCoNi and high-entropy CrMnFeCoNi alloys are comparable, the tensile strengths, tensile ductility and work of fracture of the CrCoNi alloy are significantly higher, by respectively $\sim 15$, $\sim 30$ and $\sim 50\%$, at room temperature. At cryogenic temperatures, the strengths of the two alloys are comparable ($\sigma_{UTS} \sim 1,300$ MPa at 77 K), but the $K_{JIc}$ fracture toughness, tensile ductility and work of fracture are again markedly higher in the CrCoNi alloy, by $\sim 25$, $\sim 27$ and $\sim 31\%$, respectively. The yield strength, $\sigma_y$, at 77 K is slightly below that of the CrMnFeCoNi alloy, which we believe is due to the non-uniform grain size of our present material (Fig. 1b–d). Consistent with this notion, in a previous study where the grain size of CrCoNi was uniform and comparable to that of the CrMnFeCoNi alloy, the tensile properties of the three-component alloy were found to exceed those of the five-component alloy at all temperatures[24].

**Discussion**

To seek the origins of such strength, ductility and fracture resistance between 293 and 77 K, we conducted detailed SEM analysis of the vicinity of the propagated crack; this was performed on samples sliced in two through the thickness to ensure that deformation conditions had been in fully plane strain (Fig. 3a). The EBSD scans taken in the wake of the propagated crack of a sample tested at 293 K (Fig. 3b), ahead of the crack tip of a sample tested at 198 K (Fig. 3c) and at a crack flank of a sample tested at 77 K (Fig. 3d) show grain misorientations as gradual changes in colour within individual grains indicative of significant amounts of dislocation plasticity. Similarly, back-scattered electron (BSE) scans taken on specimens fractured at room (Fig. 3b) and cryogenic temperatures (Fig. 3d) show the formation of pronounced dislocation cell structures akin to the five-component CrMnFeCoNi HEA where dislocation motion is associated with glide of ½ ⟨110⟩ dislocations on {111} planes[7], a typical deformation mechanism for fcc materials, which we presume also occurs in our three-component CrCoNi MEA. In addition, the EBSD scans show a few recrystallization twins in all samples (approximately one or two per grain) as well as the presence of deformation-induced nano-twins at 77 K (Fig. 3d). The BSE images, however, clearly reveal that deformation-induced nano-twinning is a dominant deformation mechanism occurring initially at room temperature but with increasing intensity at 198 and 77 K. From the images in Fig. 3, the nano-twins in the EBSD scans become very clear by overlaying the scan on an image quality, IQ, map of the same data set, which permits the measurement of the typical misorientation angle of 60° for twinning (Fig. 3e). We conclude from these results that between room and cryogenic temperatures where the strength, ductility and toughness of the medium-entropy CrCoNi are all simultaneously enhanced, nano-twinning contributes an important additional deformation mode that helps alleviate the deleterious effects of high strength that would normally be expected to result in lower toughness[29].

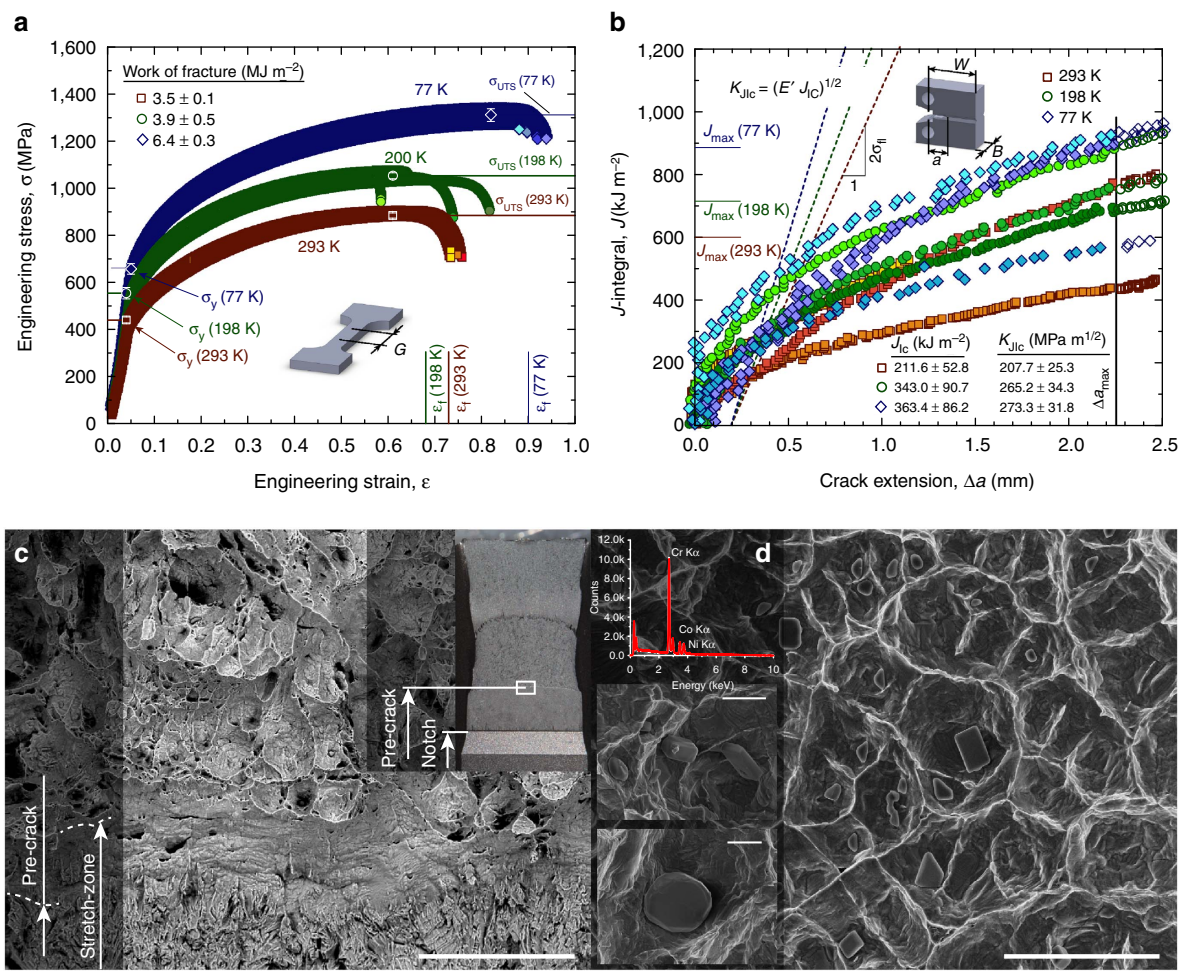

**Figure 2 | Mechanical properties and failure characteristics of the CrCoNi medium-entropy alloy.** (**a**) Tensile tests show a significant increase in yield strength, $\sigma_y$, ultimate tensile strength, $\sigma_{UTS}$ and strain to failure, $\varepsilon_f$, with decreasing temperature from room temperature, 293 K, to cryogenic temperatures, 198 and 77 K. In the same temperature range, the work of fracture increases from 3.5 MJ m$^{-2}$ to 6.4 MJ m$^{-2}$. (**b**) Fracture toughness tests on compact-tension, C(T), specimens show an increasing fracture resistance with crack extension and crack initiation, $K_{JIc}$, values of 208, 265 and 273 MPa m$^{1/2}$ at 293, 198 and 77 K, respectively. (**c**) Stereo microscopy and scanning electron microscopy images show a clear transition from the notch to the pre-crack and a pronounced stretch-zone between the pre-crack and the fully ductile fracture region of a sample that was tested at 198 K. (**d**) The fracture surface shows ductile dimpled fracture and Cr-rich particles that act as void initiation sides. (Data points shown are mean ± s.d.; see Supplementary Table 1 for exact values.) The scale bars in **c** and **d**, and the insets of **d** are 75, 5 and 2 μm, respectively.

We did not observe deformation nano-twinning at room temperature in the five-component alloy, where deformation at 293 K is solely carried by dislocation slip[7,8], specifically, involving the rapid movement of partial dislocations and the much slower planar slip of undissociated dislocations[30], although as with the present three-component alloy, twinning became a major deformation mode at 77 K. We believe that the earlier onset of deformation nano-twinning is key to the exceptional damage-tolerance of this medium-entropy alloy. Although in most materials the achievement of strength and toughness is invariably a compromise[29]—high strength is often associated with lower toughness and vice versa—it has become increasingly apparent that the presence of twinning as the dominant deformation mechanism serves to 'defeat this conflict', specifically by providing a steady source of strain hardening, which promotes ductility by delaying the onset of plastic instability by necking, and an additional deformation mode besides dislocation plasticity to accommodate the imposed strain. In addition to the high- and medium-entropy alloys, there are now several other materials known to benefit from twinning, including copper thin films[31–34] and 11–15 wt.% (Hadfield) Mn-steels (used in the mining industry for rock crushers because of their hardness and fracture resistance) and their modern variant known as twinning-induced plasticity steels[35–42], which have application in the automobile industry.

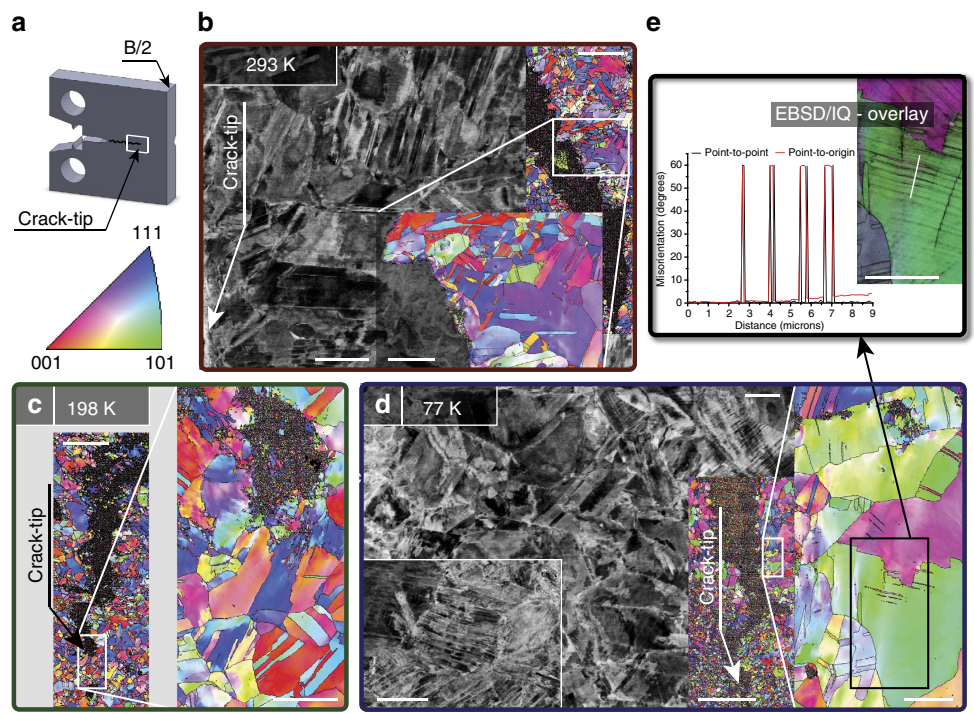

**Figure 3 | Deformation mechanisms in CrCoNi between 293 and 77 K.** (**a**) After testing, some samples were sliced in two along the half-thickness mid-plane, and the crack-tip regions in the centre of the samples (plane strain) were investigated in the scanning electron microscope using back-scattered electrons (BSE) and electron-backscatter diffraction (EBSD). (**b**) EBSD scans in the wake of the propagated crack of a sample tested at room temperature show a few recrystallization twins and grain misorientations indicative of dislocation plasticity whereas BSE scans reveal cell formation and nano-twinning as additional deformation mechanism. (**c**) Similar to room temperature behaviour, EBSD scans of samples tested at 198 K show recrystallization twins and misorientations indicative of dislocation plasticity ahead of the propagated crack-tip. (**d**) Samples tested at 77 K show pronounced nano-twinning and the formation of dislocation cells (BSE), whereas EBSD scans reveal dislocation plasticity in the form of grain misorientations, some recrystallization twins and deformation induced nano-twins. (**e**) An arbitrarily chosen path on an EBSD image overlaid on an image quality (IQ) map shows 60° misorientations typical for the character of such deformation twins. (The IQ map measures the quality of the collected EBSD patterns and is often used to visualize microstructural features.) The scale bars of the BSE image, the EBSD image and the inset of the EBSD image in **b** are 5, 75 and 25 µm, respectively; the ones of the EBSD image and its inset in **c** are 50 and 10 µm, respectively. The BSE image and its corresponding inset, and the EBSD image and its inset have scale bars of 10, 5, 200 and 15 µm, respectively. The scale bar in **e** is 15 µm.

The current medium-entropy CrCoNi alloy, however, appears to optimize these features to achieve literally unparalleled mechanical performance at low temperatures. Although solid-solution hardening provides the ideal hardening mechanism for cryogenic use, the increasing role of nano-twinning with decreasing temperature, as is evident from the apparently denser network of nano-twins at 77 K (inset in Fig. 3d) compared with room temperature (Fig. 2b), acts to progressively further enhance damage-tolerance (strength, ductility and toughness) with decreasing temperature, to achieve extremely high strain-hardening exponents on the order of 0.4.

Such damage-tolerant properties of the CrCoNi medium-entropy alloy are literally unprecedented for mechanical behaviour at cryogenic temperatures. For a material with a tensile strength of 1.3 GPa to display ductilities (failure strains) of 90%, and 'valid' crack-growth fracture toughnesses that exceeds 430 MPa m$^{1/2}$, all at liquid-nitrogen temperatures, is exceptional and clearly exceed the excellent cryogenic properties of our previously reported CrMnFeCoNi high-entropy alloy[8]. Its ductility compares favourably to high-Mn twinning-induced plasticity steels[35–42] and strength and toughness are comparable to the very best cryogenic steels, for example, certain austenitic stainless steels[43–47] and high-Ni steels[48–51]; in addition, the strength, ductility and toughness of the CrCoNi alloy exceed the properties of all medium- and high-entropy alloys reported to date (Fig. 4). Moreover, with a uniformly fine grain size, it is eminently feasible that the strength, ductility and toughness properties of this CrCoNi alloy may further improve.

With respect to high-entropy alloys in general, by comparing the CrCoNi and CrMnFeCoNi alloys, the current work does lend credence to our belief that it is the nature of elements in complex solid solutions that is more important than their mere number. Indeed, in terms of (valid) crack-initiation and crack-growth toughnesses, the CrCoNi medium-entropy alloy represents one of the toughest materials in any materials class ever reported.

## Methods

**Materials processing and microstructural characterization.** The CrCoNi MEA was produced from high-purity elements (>99.9% pure), which were arc-melted under argon atmosphere and drop-cast into rectangular cross-section copper moulds measuring 25.4 × 19.1 × 127 mm. The ingots were homogenized at 1,200 °C for 24 h in vacuum, cut in half length-wise and then cold-forged and cross-rolled at room temperature along the side that is 25.4 mm to a final thickness of ~10 mm, as shown in Fig. 1a (total reduction in thickness of ~60%). Each piece was subsequently annealed at 800 °C for 1 h in air leading to sheets with a fully recrystallized microstructure consisting of equiaxed grains ~5–50 μm in size.

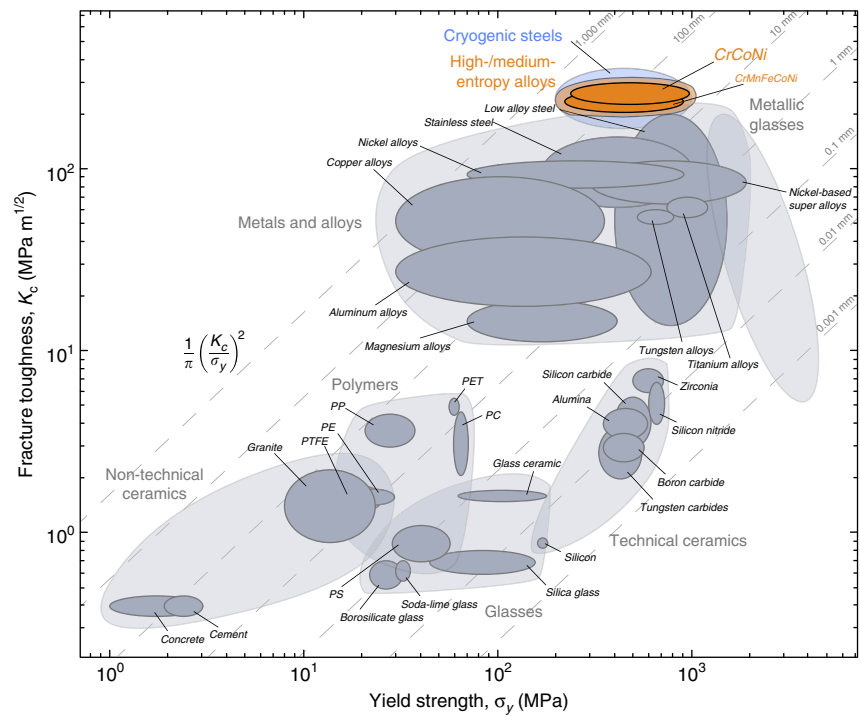

**Figure 4 | Ashby map of fracture toughness versus yield strength for various classes of materials.** The investigated medium-entropy alloy CrCoNi compares favourably with materials classes like metals and alloys and metallic glasses. Its combination of strength and toughness (that is damage tolerance) is comparable to cryogenic steels, for example, certain austenitic stainless steels[43–47] and high-Ni steels[48–51], and exceeds all high- and medium-entropy alloys reported to date.

To analyse the microstructure of the material after processing, two pieces were cut from the recrystallized sheets perpendicular to the rolling direction, embedded in conductive resin and metallographically polished in stages to a final surface finish of 0.04 μm using colloidal silica. For optical microscopy analysis, one polished surface was chemically etched using a standard solution for austenitic steels (10 ml $H_2O$, 1 ml $HNO_3$, 5 ml HCl and 1 g $FeCl_3$); the other was analysed as is in an LEO (Zeiss) 1525 FE-SEM (Carl Zeiss, Oberkochen, Germany) scanning electron microscope (SEM) operated at 20 kV in the back-scattered electron mode.

**Mechanical characterization.** Rectangular dog-bone-shaped tensile specimens with a gauge length of 12.7 mm were machined from the recrystallized sheets by electrical discharge machining. Both sides of the specimen were ground using SiC paper resulting in a final thickness of ~1.5 mm and a gauge width of ~3.0 mm. The gauge length was marked with Vickers microhardness indents (300 g load) to enable elongations to be measured after fracture using a Nikon travelling microscope. Tensile tests were performed at an engineering strain rate of $10^{-3} s^{-1}$ in a screw-driven Instron 4,204 load frame. Groups of four samples were tested at three different temperatures ($N = 12$); at room temperature (293 K), in a bath of dry ice and ethanol (198 K), and in a bath of liquid nitrogen (77 K).

The elongation of the gauge length of each sample was measured after testing, and engineering stress–strain curves were calculated from the load-displacement data. Yield strength, $\sigma_y$, ultimate tensile strength, $\sigma_u$, and elongation to failure, $\varepsilon_f$, were determined from the uniaxial tensile stress–strain curves and are shown in Supplementary Table 1 as mean ± s.d. for each set of tests at the individual temperatures. True stress–strain curves were calculated from the engineering stress–strain curves and strain-hardening exponents, $n$, were determined for each temperature based on the constitutive law $\sigma = \kappa\varepsilon^n$, where $\sigma$ and $\varepsilon$ are, respectively, the true stress and plastic strain, $k$ is a scaling constant and $n$ the strain-hardening exponent; $n$ values are also listed in Supplementary Table 1.

Nine ($N = 9$) compact-tension C(T) specimens, of nominal width $W = 18$ mm and thickness $B = 9$ mm, were prepared in strict accordance with ASTM standard E1820 (ref. 26) using electrical discharge machining (EDM). Notches, 6.6 mm in length with notch root radii of ~100 μm, were cut using EDM; before pre-cracking, the faces of all samples were metallographically ground and polished in stages to a final 1 μm surface finish to allow accurate crack-length measurements using optical microscopy. All the samples were fatigue pre-cracked and tested using an electroservo-hydraulic MTS 810 load frame (MTS Corporation, Eden Prairie, MN, USA) controlled by an Instron 8800 digital controller (Instron Corporation,

Norwood, MA, USA). Fatigue pre-cracks were created under load control (tension–tension loading) at a stress intensity range of $\Delta K = K_{max} - K_{min}$ of 15 MPa m$^{1/2}$ and a constant frequency of 10 Hz (sine wave) with a load ratio $R = 0.1$, where $R$ is the ratio of minimum to maximum applied load. During pre-cracking, the crack length was optically checked from both sides of the sample to ensure a straight crack front with crack extension monitored using an Epsilon clip gauge of 3 mm (−1/+2.5 mm) gauge length (Epsilon Technology, Jackson, WY, USA) mounted at the load-line of the sample; final crack lengths, $a$ were in the range of 8.1–12.6 mm ($a/W \sim 0.45$–0.7) and thus were well above the ASTM standard's minimum length requirement for a pre-crack of 1.3 mm. To improve the constraint conditions at the crack tip during testing, all the samples were side-grooved using EDM to depths of ~1 mm, which resulted in a net sample thickness of $B_N \sim 7$ mm; this reduction in thickness did not exceed 20–25%, as mandated by ASTM Standard E1820 (ref. 26).

Nonlinear-elastic fracture mechanics methodologies were used to incorporate both the elastic and inelastic contributions to the measurement of the fracture toughness; specifically, the change in crack resistance with crack extension, that is, crack-resistance curve (R-curve) behaviour, was characterized in terms of the J-integral as a function of crack growth at three different temperatures: 293, 198 and 77 K. The samples were tested under displacement control at a constant displacement rate of 2 mm min$^{-1}$. The onset of cracking as well as subsequent subcritical crack growth were determined by periodically unloading the sample (~20% of the peak-load) to record the elastic unloading compliance using an Epsilon clip gauge of 3 mm (−1/+7 mm) gauge length (Epsilon Technology, Jackson, WY, USA) mounted in the load-line of the sample. Crack lengths, $a_i$ were calculated from the compliance data obtained during the test using the compliance expression of a C(T) sample at the load-line[26]:

$$a_i/W = 1.000196 - 4.06319u + 11.242u^2 - 106.043u^3 + 464.335u^4 - 650.677u^5, \quad (1)$$

where

$$u = \frac{1}{\left[B_e E C_{c(i)}\right]^{1/2} + 1}. \quad (2)$$

$C_{c(i)}$ is the rotation-corrected, elastic unloading compliance and $B_e$ the effective sample thickness of a side-grooved sample calculated as $B_e = B - (B - B_N)^2/B$. (Initial and final crack lengths were additionally verified by post-test optical measurements.) For each crack length data point, $a_i$, the corresponding $J_i$-integral was computed as the sum of elastic, $J_{el\,(i)}$, and plastic components, $J_{pl\,(i)}$, such that the J-integral can be written as follows:

$$J_i = K_i^2/E' + J_{pl(i)}, \quad (3)$$

where $E' = E$, the Young's modulus, in plane stress and $E/(1 - v^2)$ in plane strain; $v$

is Poisson's ratio. $K_i$, the linear elastic stress intensity corresponding to each data point on the load-displacement curve, was calculated for the C(T) geometry from:

$$K_i = \frac{P_i}{(BB_N W)^{1/2}} f(a_i/W), \quad (4)$$

where $P_i$ is the applied load at each individual data point and $f(a_i/W)$ is a geometry-dependent function of the ratio of crack length, $a_i$, to width, $W$, as listed in the ASTM standard. The plastic component of $J_i$ can be calculated from the following equation:

$$J_{pl(i)} = \left[ J_{pl(i-1)} + \left( \frac{\eta_{pl(i-1)}}{b_{(i-1)}} \right) \frac{A_{pl(i)} - A_{pl(i-1)}}{B_N} \right] \left[ 1 - \gamma_{(i-1)} \left( \frac{a_{(i)} - a_{(i-1)}}{b_{(i-1)}} \right) \right], \quad (5)$$

where $\eta_{pl\ (i-1)} = 2 + 0.522\ b_{(i-1)}/W$ and $\gamma_{pl\ (i-1)} = 1 + 0.76\ b_{(i-1)}/W$. $A_{pl\ (i)} - A_{pl\ (i-1)}$ is the increment of plastic area underneath the load-displacement curve, and $b_i$ is the uncracked ligament width (that is, $b_i = W - a_i$). Using this formulation, the value of $J_i$ can be determined at any point along the load-displacement curve and together with the corresponding crack lengths, the $J - \Delta a$ resistance curve created. (Here, $\Delta a$ is the difference of the individual crack lengths, $a_i$, during testing and the initial crack length, $a$, after pre-cracking.)

The intersection of the resistance curve with the 0.2 mm offset/blunting line ($J = 2\ \sigma_0\Delta a$; where $\sigma_0$ is the flow stress) defines a provisional toughness $J_Q$, which can be considered as a size-independent (valid) fracture toughness, $J_{Ic}$, provided the validity requirements for $J$-field dominance and plane-strain conditions prevail, that is , that $B$, $b_0 > 10\ J_Q/\sigma_0$, where $b_0$ is the initial ligament length. The fracture toughness expressed in terms of the stress intensity was then computed using the standard $J - K$ equivalence (mode I) relationship $K_{JIc} = (E'\ J_{Ic})^{1/2}$. Values for $E$ and $\nu$ at the individual temperatures were determined by resonance ultrasound spectroscopy using the procedure described in Haglund et al.[27]; at 293, 198 and 77 K, Young's moduli, $E$ of 229, 235 and 241 GPa and Poisson's ratios, $\nu$ of 0.31, 0.30 and 0.30 were used, respectively.

To discern the mechanisms underlying the measured fracture toughness values and investigate the microstructure in the vicinity of the crack tip and wake in the plane-strain region in the interior of the sample after testing, one sample from each of the tested temperatures was sliced in two, each with a thickness of $\sim B/2$. For each sample, one half was embedded in conductive resin, progressively polished to a 0.04 μm surface finish using colloidal silica, and analysed in the SEM in back-scattered electron mode as well as by electron back-scatter diffraction, EBSD using a TEAM EDAX analysis system (Ametek EDAX, Mahwah, NJ, USA).

The remaining ligament of all other samples was cycled to failure at a $\Delta K$ of $\sim 30\ \text{MPa m}^{1/2}$, a frequency of 100 Hz (sine wave) and a load ratio $R = 0.5$ so that both the initial and the final crack lengths could be optically determined with precision from the change in fracture mode. In addition, the mating fracture surfaces of each sample were examined in the SEM at an accelerating voltage of 20 kV in the secondary electron mode. Particles inside the microvoids of samples tested at 293 and 77 K were analysed using an Energy Dispersive Spectroscopy (EDS) system from Oxford Instruments (Model 7426, Oxford, England). EDS analyses were performed on five randomly chosen particles from samples tested at both room and liquid nitrogen temperature to determine their chemical composition.

### Acknowledgements
This research was sponsored by the U.S. Department of Energy, Office of Science, Office of Basic Energy Sciences, Materials Sciences and Engineering Division, through the Materials Science and Technology Division at the Oak Ridge National Laboratory (for H.B, Z.W. and E.P.G.) and the Mechanical Behavior of Materials Program (KC13) at the Lawrence Berkeley National Laboratory (for B.G., K.V.S.T. and R.O.R.).

### Author contributions
B.G., E.P.G. and R.O.R. designed the research; H.B. and E.P.G. made the alloy; B.G., A.H., K.V.S.T., H.B. and Z.W. mechanically characterized the alloy; B.G., A.H., K.V.S.T., H.B., E.P.G. and R.O.R. analysed and interpreted the data; B.G., E.P.G. and R.O.R. wrote the manuscript.

### Additional information
**Supplementary Information** accompanies this paper at http://www.nature.com/naturecommunications

**Competing financial interests:** The authors declare no competing financial interests.

**Reprints and permission** information is available online at http://npg.nature.com/reprintsandpermissions/

**How to cite this article:** Gludovatz, B. *et al.* Exceptional damage-tolerance of a medium-entropy alloy CrCoNi at cryogenic temperatures. *Nat. Commun.* 7:10602 doi: 10.1038/ncomms10602 (2016).